\documentstyle[aps,prl,multicol]{revtex}

\def\date#1{\author{\small(#1)}}
\def\abstract#1{\author{\parbox[t]{5.5in}{\small#1}}\par\maketitle}

\def\f.#1.{{\bf #1}}
\def\mb.#1.{\bbox{#1}}

\def\opn{\begin{equation}} \def\cls{\end{equation}}
\def\opa{\begin{eqnarray}} \def\cla{\end{eqnarray}}
\def\opbib{\begin{references}} \def\clbib{\end{references}}
\def\bb#1{\bibitem{#1}}
\def\qno;#1;{\label{#1}\end{equation}} \def\qna;#1;{\label{#1}\end{eqnarray}}
\def\rf;#1;{(\ref{#1})}
\def\dels#1 {\nabla\kern -1.5pt_{#1}\kern 1.5pt}
 \def\av#1{\left\langle #1\right\rangle}

\def\suspend{\end{multicols}\vspace*{-0.5cm} \noindent \rule{8.65cm}{.02cm}}
\def\resume{\hskip 9.3cm \rule{8.65cm}{.02cm} 
\begin{multicols}{2}\vspace*{-0.8cm} \noindent}
\def\umo{\rlap{\"{$\,$\ }} \,\,\,\kern -.3 cm o}

\newcount\hours \newcount\minutes \newcount\a \newcount\b
\def\gmt{\hours = \time \divide\hours by 60 \a =\hours \multiply \a by 60
\minutes = \time \advance \minutes by -\a
{\ifnum\hours<10 0\fi}\number\hours
{\ifnum\minutes<10 0\fi}\number\minutes}
\def\gday{\b=\year \advance \b by -1900
\number\b{\ifnum\month<10 0\fi}\number\month{\ifnum\day<10 0\fi}\number\day}
\def\today{\number\day\ 
\ifcase\month\or January\or February\or March\or April\or May\or 
June\or July\or August\or September\or October\or November\or December\fi
\ \number\year}

\def\al{\alpha} \def\be{\beta}  \def\de{\delta}
    
  \def\rh{\rho}

\def\casefr#1/#2 {\case{#1}{#2}}
\def\part{\partial}

  \def\prop{\propto}

\def\sump#1{\lower .15in\hbox{$\stackrel{\displaystyle{\sum}'} {\scriptstyle 
#1}$}}

\def\tZ{\tilde Z}

\begin{document}

\title{\vbox to 0pt {\vskip -1cm \rlap{\hbox to \textwidth {\rm{\small To 
appear in Phys. Fluids\hfill} }}}Note on Forced Burgers Turbulence}

\author{Robert H. Kraichnan$^*$}

\address{PMB 108, 369 Montezuma, Santa Fe, NM 87501-2626}

\date{30 July 1999}

\abstract{A putative powerlaw range of the probability density of velocity 
gradient in high-Reynolds-number forced Burgers turbulence is studied. In the 
absence of information about shock locations, elementary conservation and 
stationarity relations imply that the exponent $-\al$ in this range satisfies 
$\al\ge3$, if dissipation within the powerlaw range is due to isolated shocks. 
A generalized model of shock birth and growth implies $\al=7/2$ if initial 
data and forcing are spatially homogeneous and obey Gaussian statistics. 
Arbitrary values $\al\ge3$ can be realized by suitably constructed 
homogeneous, non-Gaussian initial data and forcing.}

\vskip .4cm

\begin{multicols}{2}

Burgers equation was originally proposed as a simplified dynamical system 
that might point at statistical procedures applicable to Navier-Stokes 
turbulence. That goal seems still far away. Meanwhile, a body of research has 
been devoted to Burgers turbulence itself. The particular topic addressed in 
the present paper is the power law of a part of a probability density for 
Burgers turbulence forced at high Reynolds number. The conclusion reached is 
that determination of the correct power law requires detailed examination of 
the dynamical behavior of explicit structures, namely shocks. This perhaps has 
discouraging implications for any general theory of the higher statistics of 
Navier-Stokes turbulence, where more varied and plastic flow structures must 
be confronted.

There has been considerable interest in a putative powerlaw range of the 
probability density $Q(\xi)$ of velocity gradient $\xi=u_x$ for 
high-Reynolds-number Burgers turbulence forced at large scales. This range, of 
form $Q(\xi) \prop |\xi|^{-\al}$, is believed to occupy negative $\xi$ values 
intermediate between those near the central peak of $Q$ and those 
characteristic of the shock interiors. Proposals for $\al$ include $2$ 
\cite{1,2}, the range $5/2$ to $3$ \cite{3,4}, $3$ \cite{5}, and $7/2$ 
\cite{6,7,8,9}.

It should be emphasized at the outset that the high-Reynolds-number limit is 
not the only case of theoretical interest. As with Navier-Stokes turbulence, 
the construction of faithful analytical approximations at finite Reynolds 
numbers remains a challenge.

E and Vanden Eijnden \cite{7,8,9} advance and clarify the mathematics of the 
infinite-Reynolds-number limit and provide some valuable tools. One is a 
simple representation of the effects of shock interactions on $Q$ in terms of 
the rate at which fluid is swallowed by the shocks. Another is a steady-state 
integral representation of the asymptotic large-$|\xi|$ form of $Q(\xi)$ in 
terms of the dissipation term $F(\xi)$ in the $Q$ equation of motion. Another 
is an explicit shock-birth model that implies $\al=7/2$.

The present paper explores the relation between the form of $F(\xi)$ and the 
value of $\al$. If viscous effects in the $-\al$ range are due to isolated 
shocks, the form of $F(\xi)$ in this range expresses the relative likelihood, 
weighted by shock strength, that a shock occurs in a fluid environment with 
given $\xi$.

In the absence of an explicit shock-growth model, or other source of 
information about the distribution of shocks, the integral equation for $Q$ is 
found to yield $\al \ge 3$. The slightly stronger bound $\al > 3$ is stated in 
\cite{7} and \cite{8}, but with the recognition that $\al=3$ may be realized 
under particular circumstances. The limit of infinite Reynolds number is taken 
in the present paper without making a split \cite{7} of the velocity field 
into shock interiors and external field. At the end, the analysis is extended 
to the split-field representation.

A more general model of shock growth is presented here. It is independent of 
details of internal shock structure. In this model, one examines the length of 
time during which $\xi$ can steepen within a fluid element before the fluid 
element hits a shock. The model implies $\al=7/2$ if the forcing is 
statistically homogeneous and Gaussian. More general statistically homogeneous 
forcing can realize arbitrary values $\al\ge3$.

Let $R=u_0L/\nu$, $\xi_0=u_0/L$, $\xi_S=R\xi_0$, where $u_0$, $L$, $\nu$ are 
root-mean-square velocity, spatial macroscale, and viscosity. The order of 
magnitude of gradients within typical shocks is $\xi_S$. The forced Burgers 
equation is
\opn
u_t + u_xu = \nu u_{xx} + f,
\qno;1;
where $f$ is the (statistically stationary) forcing field. If $f$ has 
infinitely short correlation times, \rf;1; leads to
\opn
Q_t = \xi Q + (\xi^2Q)_\xi + BQ_{\xi\xi} + F,
\qno;2;
where
\opn
F(\xi,t) = -\nu(H(\xi,t)Q(\xi,t))_\xi, \quad H(\xi,t) = \av{\xi_{xx}|\xi},
\qno;3;
and the parameter $B$ measures the strength of forcing of $\xi$ \cite{5}. The 
first term on the right side of \rf;2; represents loss or gain of measure due 
to squeezing or stretching of the fluid, the second term describes relaxation 
of positive $\xi$ and steepening of negative $\xi$, and $F$ includes all 
viscous effects.

Statistical homogeneity requires $\av{\xi} = \int_{-\infty}^\infty \xi Qd\xi 
= 0$. Multiplication of \rf;2; by $\xi$ and integration over all $\xi$ shows 
that this condition is preserved, provided that $Q$ vanishes strongly enough 
at $\pm\infty$. The result depends on
\opn
\int_{-\infty}^{\infty} \xi F(\xi)d\xi = 0,
\qno;4;
which follows from \rf;3; and $\av{\xi_{xx}}=0$.

If $f$ has spectral support effectively confined to wavenumbers $O(1/L)$, 
then $\xi_0 = O(B^{1/3})$. In this case, it is widely agreed that the 
steady-state $Q$ has a complicated form in the limit $R\to\infty$. There is a 
central peak of width $O(\xi_0)$ whose form is $R$-independent in the limit. 
There is faster-than-algebraic decay as $|\xi|\to\infty$. For $\xi > 0$, this 
decay has the specific form $\exp(-\xi^3/3B)$, with an algebraic prefactor. 
The rapidly-decaying tail for $\xi < 0$ (far-left tail) includes $|\xi| \ge 
O(\xi_S)$. It is preceded, at smaller $|\xi|$, by an algebraic tail of form 
$1/R|\xi|$ ($-1$ range) associated with the shoulders of developed shocks. 
Between the $-1$ range and the central peak, an inner algebraic tail of form 
$Q(\xi) \prop \xi_0^{\al-1}|\xi|^{-\al}$ is expected. This tail is driven by 
the inviscid steepening of negative gradients. Proposals for the value of 
$\al$ have ranged from $2$ to $7/2$.

The $-\al$ range is infinite if $R$ is, but is confined to $|\xi|$ smaller 
than the $O(\xi_S)$ gradients inside the shocks. Thus the range is restricted 
to $\xi > -\xi_M \equiv -R^z\xi_0$, with $z \le 1$. In fact, both shock 
analysis and simulation show that the $-\al$ range is masked by the $-1$ range 
at negative enough $\xi$. The masking further restricts the observable $-\al$ 
range to $z = 1/(\al-1)$, a result that follows from setting $1/R\xi_M = 
\xi_0^{\al-1}\xi_M^{-\al}$. It should be emphasized that the transition from 
$Q(\xi) \prop |\xi|^{-\al}$ to $Q(\xi) \prop 1/R|\xi|$ at $\xi = O(-\xi_M)$ is 
a masking, not a dynamical transition. The $-1$ range is associated with the 
shoulders of quasi-stationary mature shocks while the $-\al$ range is 
associated with inviscid steepening of gradients away from shocks. The latter 
process continues, for at least some fluid elements, up to $|\xi| = O(\xi_S)$.

The behavior of $Q$ in all ranges is linked by \rf;2; to the form of 
$F(\xi)$. For large negative $\xi$, the relationship is expressed in 
especially clear form by the integral representation
\opn
Q(\xi) \approx |\xi|^{-3}\int_{-\infty}^\xi \xi'F(\xi')d\xi' \quad (-\xi \gg 
\xi_0)
\qno;5;
derived by E and Vanden Eijnden \cite{7}. An alternative form of \rf;5; is 
\cite{5}
\opn
\nu H(\xi) \approx \xi^2 + {1\over Q(\xi)}\int_{-\infty}^\xi Q(\xi')\xi'd\xi' 
\quad (-\xi \gg \xi_0).
\qno;6;
or by \rf;3;,
\opn
F \approx -\xi Q - (\xi^2Q)_\xi \quad (-\xi \gg \xi_0).
\qno;7;
Equation \rf;7; says that in a steady state, for $\xi$ such that the $B$ term 
is negligible, the $F$ term in \rf;2; must balance both the gradient 
steepening term $(\xi^2Q)_\xi$ and the loss-of-measure term $\xi Q$. Equations 
\rf;5;--\rf;7; do not require large $R$.

According to \rf;5;, the behavior of $Q$ in the $-\al$ range depends on, 
first, the value $\int_{-\infty}^{-\xi_M}\xi'F(\xi')d\xi'$ of the integral at 
the outer end of the range; second, the $\xi$ dependence of $F(\xi)$ within 
the $-\al$ range, which determines the growth of the integral within the 
range. If $F(\xi)$ in the $-\al$ range arises solely from isolated shocks, the 
analysis of \cite{7} gives an immediate bound on $\al$. This analysis 
expresses a simple physics: If a shock lies in surrounding fluid with gradient 
$\xi$, then this fluid is swallowed by the convergence at the shock, thereby 
decreasing $Q(\xi)$. This means that $F(\xi)$ cannot be positive in the $-\al$ 
range. Since $\xi < 0$, the integral in \rf;5; therefore cannot increase as 
$|\xi|$ decreases within the range, and $\al < 3$ is impossible.

The cases $\al=3$ and $\al>3$ imply qualitatively different magnitudes in 
\rf;5;. If $\al=3$, then \rf;5; immediately gives 
$\int_{-\infty}^{-\xi_M}\xi'F(\xi')d\xi' = O(\xi_0^2)$. It is then required 
that $F$ be small enough that $\int_{-\infty}^\xi\xi'F(\xi')d\xi' = 
O(\xi_0^2)$ if $\xi$ lies anywhere within the $-\al$ range. If $\al > 3$, 
\rf;5; immediately gives $\int_{-\infty}^{-\xi_M}\xi'F(\xi')d\xi' \to 0$ as 
$\xi_M/\xi_0\to\infty$. Then $F(\xi) \prop -|\xi|^{1-\al}$ within the range 
will give $Q(\xi) \prop |\xi|^{-\al}$. This means $F(\xi) \prop \xi Q(\xi)$.

If dissipation within the $-\al$ range is due to isolated shocks, the 
physical difference between $\al=3$ and $\al>3$ lies in the weighted relative 
likelihood that shocks exist in environments with given $\xi$. $F(\xi) \prop 
-Q(\xi)$, a case of Polyakov's closure \cite{3,4}, is one form that is 
consistent with $\al=3$. This corresponds to unbiased placement of shocks in 
the fluid \cite{10}. $F(\xi) \prop \xi Q(\xi)$ represents bias toward larger 
$|\xi|$ within the $-\al$ range. Since $Q(\xi)$ falls off rapidly in the 
$-\al$ range, even this form puts most of the shocks in fluid with $|\xi| = 
O(\xi_0)$.

If $\al=3$, the inviscid steepening of gradient in most fluid elements with 
negative $\xi$ continues until $|\xi| = O(\xi_S)$. If the dissipative process 
that stops further steepening is absorption by well-defined shocks, these 
shocks must occur with sufficient weight in environments with 
$|\xi|=O(\xi_S)$. Since $\xi_S$ measures typical gradients within shocks, 
other processes, such as new shock formation, may actually dominate the 
dissipation.

Further insight concerning the relation between $F$ and $\al$ is provided by 
the equation of motion \rf;2; for $Q$ and, in particular, by the steady-state 
equation $Q_t=0$. A variety of global assumed forms for $F$, including $F(\xi) 
\prop -Q(\xi)$, can be adjusted, by tuning of the constant of proportionality, 
to give a steady-state $Q(\xi)$ that vanishes strongly as $\xi\to+\infty$, is 
positive everywhere, and has a range $\al = 3$. In order to be physically 
relevant, $F(\xi)$ for negative $\xi$ beyond $\xi_M$ must be shaped to give 
consistent $1/R|\xi|$ and far-tail ranges for $Q$. Global model forms can also 
be constructed that yield a range with $\al > 3$.

The quantity $\av{\xi}$ is identically conserved at value zero. It is of 
interest to examine the way in which flow of $\av{\xi}$ through the $-\al$ 
range depends on the value of $\al$. Let \rf;2; be multiplied by $\xi$ and 
integrated over the range $(-\xi_M,\infty)$ to yield
\opn
\int_{-\xi_M}^\infty \xi Q_t(\xi)d\xi = \xi_M^3Q(-\xi_M) + 
\int_{-\xi_M}^\infty \xi F(\xi)d\xi.
\qno;8;
In writing \rf;8;, a partial integration is performed, it is assumed that $Q$ 
vanishes strongly enough at $+\infty$, and it is assumed that $\xi_M/\xi_0$ is 
large enough that the $B$ term is negligible.

The left side of \rf;8; is the rate of increase of $\av{\xi}$ in the range 
$(-\xi_M,\infty)$. On the right side, $\xi_M^3Q(-\xi_M)$ is the rate of 
increase due to flow of negative contribution to $\av{\xi}$, through the 
boundary at $-\xi_M$, to the range $(-\infty,-\xi_M)$. This flow is due to 
inviscid steepening of gradients. The $F$ term on the right side of \rf;8; is 
the rate of increase of $\av{\xi}$, or decrease of $\av{-\xi}$, in the range 
$(-\xi_M,\infty)$, due to viscous interaction with isolated shocks and any 
other dissipative structures that may be present.

The nature of the $F$ term in \rf;8; must be understood clearly. The present 
analysis is at finite $R$, with the eventual limit $R\to\infty$ considered. 
There is no sudden jump of $\av{\xi}$ contribution from $-\al$ range to shock 
interiors. What does happen at large $R$ is that a fluid element with $\xi$ in 
the $-\al$ range hits the shock and then suffers a very rapid, but continuous, 
steepening until its $\xi$ is the order of that in the shock interior. The sum 
of these events in the entire range is expressed by the $F$ term in \rf;8;.

In a steady state, the right side of \rf;8; vanishes. The implications differ 
in the cases $\al=3$ and $\al>3$. If $\al>3$, the boundary-flow term in \rf;8; 
vanishes in the limit $\xi_M/\xi_0\to\infty$. This means that the $F$ integral 
term vanishes also in the limit. In view of \rf;4;, where the limits are true 
infinity (infinite compared to $\xi_S$), it follows that 
$\int_{-\infty}^{-\xi_M} \xi F(\xi)d\xi$ also vanishes in the limit. This does 
not mean that $F(\xi)$ tends to zero in the limit for all $\xi$ in the range 
$(-\infty,-\xi_M)$. Instead, there are both positive and negative 
contributions that cancel in the limit. In general, $F(\xi)$ is positive in 
the $-1$ range, as illustrated by \rf;9; and \rf;10; below.

If $\al=3$, the boundary term in \rf;8; tends to a nonzero positive constant 
in the limit. A steady state then implies that the $F$ integral in \rf;8; is 
negative and that $\int_{-\infty}^{-\xi_M} \xi F(\xi)d\xi$ is positive. This 
means that the contributions from the negative and positive regions of $F$ in 
the range $(-\infty,-\xi_M)$ do not cancel. The nonzero value of 
$\int_{-\infty}^{-\xi_M} \xi F(\xi)d\xi$ needed in the case $\al=3$ has 
already been noted in the discussion of \rf;5;.

For both $\al>3$ and $\al=3$, the boundary flow through an arbitrary point 
$-\xi_\al$ fully within the $-\al$ range is independent of $\xi_\al$ in the 
limit $R\to\infty$. Thus if $\xi_M$ in \rf;8; is replaced by any $\xi_\al$ 
such that $\xi_\al/\xi_M \to 0$, $\xi_\al/\xi_0 \to \infty$ in the limit, then 
the limiting value of the boundary flow vanishes if $\al>3$. The flow has a 
value  $O(\xi_0^2)$, independent of $\xi_\al$, if $\al=3$.

The form of $Q$ and $F$ in the far-left tail and $-1$ range actually is not 
constrained by whether $\int_{-\infty}^{-\xi_M} \xi F(\xi)d\xi$ vanishes or is 
$O(\xi_0^2)$ as $R\to\infty$. The key is the value of $z=1/(\al-1)$, which 
gives the position of the join between $-1$ and $-\al$ ranges at 
$-\xi_M=-R^z\xi_0$. Consider the generic form
\opn
Q(\xi) \approx Z(\xi/R\xi_0)/R|\xi| \quad (-\xi \gg \xi_0)
\qno;9;
in the far-tail and $-1$ ranges, where $Z$ vanishes strongly as 
$\xi\to-\infty$, $Z \to C_z$ as $\xi/R\xi_0 \to -0$, and $C_z$ is an $O(1)$ 
constant. If this form is substituted into \rf;7;, a consequence is
\opn
\int_{-\infty}^{-\xi_M} \xi F(\xi)d\xi \approx C_zR^\be\xi_0^2 \quad (R \gg 
1),
\qno;10;
where $\be=(3-\al)/(\al-1)$. If $\al=3$, this gives the needed $O(\xi_0^2)$ 
result. If $\al > 3$, the right side vanishes at $R=\infty$. Thus the value of 
the boundary flow automatically adjusts to the value of $\al$ that is 
determined by the form of $F$ in the $-\al$ range. In other words, it adjusts 
to the probability distribution of shock occurence in the $-\al$ range.

Explicit models of shock development lead to explicit forms of $F(\xi)$. The 
following model generalizes those of \cite{6,7,8,9}. It assumes large $R$ but 
does not invoke the internal shock structure. The development is followed 
before and after shock birth. Take the unforced case first. Consider an 
initial velocity field of form
\opn
u(x,0) \approx \xi_0(-ax + bx|x/L|^p) \quad (p > 0)
\qno;11;     
in the vicinity of a point $x=0$ where a shock will form at time $1/a\xi_0$. 
Here $a$ and $b$ are $O(1)$ positive constants. If $p$ is an even integer, all 
$x$ derivatives of $u$ exist at $x=0$. If $p$ is an odd integer or 
non-integer, only the derivatives of order $n \le p+1$ exist.

This form of initial velocity field leads to $\al = 3+1/p$, a result that can 
be verified in several ways. The following is a  simple qualitative argument. 
The inviscid evolution of velocity gradient in a Lagrangian frame satisfies 
$\dot\xi=-\xi^2$. It then follows from \rf;11; that the time of initial shock 
formation is $t_0=1/a\xi_0$. The negative gradient within a fluid element 
initially at small enough $|x| > 0$ grows inviscidly until the fluid element 
hits the shock.

The time at which a fluid element initially at $x$ falls into the growing 
shock is $t_x \approx 1/[\xi_0(a-b|x/L|^p)]$. The times of arrival at the 
shock determine the fractional measure of initial points $x$ such that, before 
a fluid element hits the shock, the gradient magnitude increases to values 
$\ge |\xi| \gg \xi_0$. The gradient magnitude at $x$ grows to $\approx 
a\xi_0|x/L|^{-p}/[b(p+1)]$ at $t_0$ and $\approx a\xi_0|x/L|^{-p}/bp$ at 
$t_x$.  Thus the measure of points $x$ such that the gradient magnitude in the 
fluid element initially at $x$ can grow to a value that equals or exceeds 
$|\xi| \gg \xi_0$ within the intervals $(0,t_0)$ or $(0,t_x)$ is $\prop 
|\xi|^{-1/p}$. Of the initial fluid elements that achieve a value at least 
$|\xi| \gg \xi_0$ before hitting the shock, the fraction that does this in the 
preshock interval $(0,t_0)$ is $(p/p+1)^{1/p}$.

When the fluid element has achieved the value $\xi$, squeezing has decreased 
its measure by a factor $\prop |\xi|^{-1}$. Finally, the residence time of the 
fluid element in the gradient interval $d\xi$ at $\xi$ is $dt=|\xi|^{-2}d\xi$. 
Putting these factors together, one obtains $\bar Q(\xi) \prop 
|\xi|^{-1-2-1/p} = |\xi|^{-3-1/p}$, where $\bar Q(\xi)$ is the mean of 
$Q(\xi,t)$ over a time interval (say $2/a\xi_0$) long enough for all fluid 
elements that can achieve $|\xi| \gg \xi_0$ to have hit the shock.

In \cite{5}, $\al = 3$ was deduced under the assumption that the fractional 
measure of fluid elements that can achieve gradient magnitudes $\gg \xi_0$ is 
$O(1)$. The measure $\prop |\xi|^{-1/p}$ found instead in the present model 
changes the result to $\al=3+1/p$. For all finite $p>0$, the form of $F$ 
calculated from the present model is $F(\xi) \prop \xi Q(\xi)$, within the 
$-\al$ range.

Initial sawtooth profiles, where $u(x)$ consists solely of straight-line 
segments, correspond to $b=0$ (or $p=\infty$) in the model. They evolve into 
shocks that have finite amplitude at birth and yield $\al=3$. The present 
model thereby is consistent with the conclusion \cite{7,8,9} that isolated 
shocks can induce $\al=3$ only if they are created with finite amplitude.

The steady state produced by spatially smooth Gaussian forcing supported by 
wavenumbers $O(1/L)$ can be interpreted in terms of this model. Such forcing 
induces smooth profiles corresponding to $p=2$ near points of extremal slope. 
In the absence of force, a shock forms from steepening of slope at a point of 
maximally negative slope. Smooth change of the velocity field due to Gaussian 
forcing can change the location of the point of maximally negative slope as a 
function of time. But such forcing does not change the nature of the shock 
formation phenomenon because the quadratic decrease of slope magnitude away 
from the point of negative maximum survives. Once the slope magnitude at 
negative maximum is large compared to $\xi_0$, the forcing should have no 
significant effect on either the progression to shock birth or the initial 
shock growth after birth. The value $\al=7/2$ corresponding to $p=2$ is the 
steady-state result.

Forcing that supports the general case $p\ne2$ in a statistically steady 
state can be constructed as follows: Let the forcing consist of a set of 
$\de$-functions in time, spaced at time intervals $O(1/\xi_0)$. Let each such 
$\de$-function create an increment to $u(x)$ that consists of straight-line 
segments of length $O(L)$ with $O(u_0/L)$ positive slope, smoothly joined to 
interposed negative-slope regions of $O(L)$ or shorter lengths. In each 
negative-slope region let there be a point of maximally negative slope 
surrounded by a neighborhood in which $u(x)$ has the form 
$\xi_0(-ay+by|y/L|^p)$, where $y$ is the distance from the point of maximally 
negative slope. The values of $a$ and $b$ can change stochastically from one 
such region to another. Under these conditions, the negative-slope increment 
to $u(x)$ created by each $\de$-function force field is added to an existing 
field that has locally constant slope with $O(1)$ probability. Therefore the 
points of maximally-negative slope are at the special points $y=0$, and the 
consequent shock development yields $\al=3+1/p$ for all $p > 0$, as in the 
initial-value case.

All cases of the model except $p=2$, $\al=7/2$ require special shapes of the 
velocity field prior to shock formation and, therefore, precise phase 
relations of spatial Fourier components. If the forcing field is spatially 
homogeneous and has an infinitely short coherence time (white forcing), the 
effective forcing is Gaussian and these shapes and phase relations cannot be 
realized. The white, homogeneous forcing assumed in writing the $B$ term in 
\rf;2; implies $p=2$, $\al=7/2$.

At the end of \cite{5}, it was noted that the value of $\al$ depends on how 
likely it is for a shock collision to interrupt the steepening of negative 
gradients. It was argued that collisions should be infrequent enough that 
inviscid steepening should survive for most fluid elements of negative $\xi$ 
until $|\xi| = O(\xi_S)$ is reached. This implies $\al=3$. The present 
shock-growth model, following the earlier ones in \cite{6,7}, says instead 
that fluid elements with large negative $\xi$ are inevitably close to shocks 
that soon swallow them if $p=O(1)$. For most fluid elements with negative 
$\xi$, the inviscid steepening is terminated by collision with the shocks 
while $|\xi|$ is still $O(\xi_0)$.

The analysis in \cite{7} is done in terms of a split of $u$ and $\xi$ into 
parts exterior to shocks and parts interior to shocks, in the limit 
$R\to\infty$:
\opn
u(x,t) = u_e(x,t) + u_i(x,t), \quad \xi(x,t) = \xi_e(x,t) + \xi_i(x,t)
\qno;12;
It is of interest to discuss how to make the split \rf;12; when $R$ is large 
but not infinite, and to express the preceding analysis in terms of the 
split-field representation. The split into interior and exterior fields is  
analyzed in \cite{8,9} by means of matched asymptotic expansions.

If the $R=\infty$ field consists of infinitely thin shocks surrounded by 
fluid in which $|\xi| < \xi_1$, where $\xi_1$ is some finite bound, the split 
is clear and unambiguous. At large finite $R$, one can isolate a small region 
surrounding each shock \cite{7} as $u_i$, and consistently let the widths of 
these regions shrink to zero at $R=\infty$. E and Vanden Eijnden show that the 
result is a simple set of statistical equations relating $\xi_e$ and the shock 
jumps:
\opn
\av{\xi_e} + \rh\av{s} = 0,
\qno;13;
\opn
F_e(\xi_e,t) = {\rh\over2}\int s\left[V_-(\xi_e,s,t) + V_+(\xi_e,s,t) 
\right]ds.
\qno;14;
Here $\rh$ is the number density of shocks, $s$ is shock-jump strength 
($<0$), $F_e$ is the viscous term in an equation of motion like \rf;2; for the 
probability density $Q_e(\xi_e)$ of the exterior field, and $V_-$ ($V_+$) is 
the probability that a shock of strength $s$ has a left (right) environment 
with gradient $\xi_e$.

These equations have a direct physical interpretation. Equation \rf;14; 
expresses the viscous term as the rate at which $Q_e(\xi_e)$ is diminished 
through  the swallowing of fluid with gradient $\xi_e$ by shocks. The time 
derivative of \rf;13; is an expression of the fact that the rate of change of 
shock strength is given by the product of the convergence velocity and the 
negative of the gradient of the external field, as the latter is swallowed by 
the shock.

The split into $\xi_e$ and $\xi_i$ is less clear if there is a $-\al$ range 
at $R=\infty$ that includes $|\xi|$ values larger than any finite bound. Then 
the $\xi_e$ field is not smooth. As $R$ approaches infinity, it is not 
possible to form an $R$-dependent parameter $\xi_1$ such that all field with 
$|\xi| < \xi_1$ belongs to $\xi_e$ and all field with $|\xi| > \xi_1$ belongs 
to $\xi_i$. Instead the division into interior and exterior fields must be 
made individually at each shock.

The equation of motion for $\xi$ obtained by differentiation of \rf;1; 
contains the steepening term $\xi^2$ and the dissipation term $\nu\xi_{xx}$. 
One possibility for splitting the $\xi$ field is to let $\xi_e$ include all 
points $\xi>0$ and all points $\xi<0$ that are exterior to defined boundaries 
of the shock shoulders. A shock-shoulder boundary could be defined as a point 
where, as one moves through the field toward the shock, $\xi^2/|\nu\xi_{xx}|$ 
first falls below some prescribed ratio, say $10$. Then $\xi_i$ constitutes 
the field interior to the left and right shock-shoulder boundaries. This gives 
a $\xi_e$ field that extends up to values $|\xi| = O(\xi_S)$ at some points, 
with a $Q_e(\xi_e)$ that must deviate from powerlaw behavior for such $|\xi|$ 
values. The qualitative behaviors of $\xi_e$ and $\xi_i$ are independent of 
the precise value prescribed for the critical ratio of steepening term to 
dissipation term.

Since $|\xi_e|$ extends to $\infty$ in the limit, it is not obvious that all 
interactions between exterior and interior fields have a form consistent with 
\rf;13; and \rf;14;. In the limit, the $1/R|\xi|$ range belongs to $\xi_i$, 
which includes the shock shoulders. Both $\xi_e$ and $\xi_i$ contribute to the 
total-field probability density $Q(\xi_M)$, and this fact remains as 
$R\to\infty$.

It is been remarked above that $F(\xi)$ is positive in the $-1$ range, while 
\rf;14; gives negative $F_e(\xi_e)$. This emphasises that the region of $\xi$ 
contributing to the $-1$ range in each individual shock boundary layer should 
be assigned to $\xi_i$.

Equations similar to \rf;2;, \rf;5;, and \rf;8; can be written for 
$Q_e(\xi_e)$:
\opn
(Q_e)_t = \xi_e Q_e + (\xi_e^2Q_e)_\xi + B(Q_e)_{\xi\xi} + F_e,
\qno;15;
\opn
Q_e(\xi_e) \approx |\xi_e|^{-3}\int_{-\infty}^{\xi_e} 
\xi_e'F_e(\xi_e')d\xi_e' \quad (-\xi_e \gg \xi_0),
\qno;16;
\opn
\int_{-\xi_\al}^\infty \xi_e (Q_e)_t(\xi_e)d\xi_e = \xi_\al^3Q_e(-\xi_\al) + 
\int_{-\xi_\al}^\infty \xi_e F_e(\xi_e)d\xi_e.
\qno;17;
Although the equations look the same, $F_e$ and $F$ behave differently for 
negative arguments.

The parameter $\xi_M$ has no special significance here because the $-\al$ 
range of $Q_e$ is not masked by the $-1$ range. The latter belongs to $\xi_i$. 
Therefore $\xi_M$ has been replaced in \rf;17; by $\xi_\al$, which is any 
value within the $-\al$ range that satisfies the limiting relations
\opn
\xi_\al/\xi_0 \to \infty, \quad \xi_\al/\xi_S \to 0 \quad (R \to \infty)
\qno;18;
If $\al=3$, \rf;17;, like \rf;8;, exhibits a boundary flow that does not 
vanish at $R=\infty$.

The analog of \rf;4;, $\int_{-\infty}^\infty \xi_eF_e(\xi_e)d\xi_e = 0$, 
holds in steady states. In general transient states, \rf;13; implies that 
there is an additional term involving the rate of change of shock jumps.

The power-law behavior of $Q_e(\xi_e)$ must change to a faster (eventually 
faster-than-algebraic) fall-off for $|\xi_e| \ge O(\xi_S)$. The generic form 
may be written
\opn
Q_e(\xi_e) \approx \xi_0^{\al-1}|\xi_e|^{-\al}\tZ(\xi_e/R\xi_0) \quad (-\xi_e 
\gg \xi_0).
\qno;19;
Here $\tZ$ vanishes strongly at $\xi_e=-\infty$ and is $O(1)$ at $\xi_e=-0$. 
The precise form of $\tZ$ is $\al$-dependent. In contrast to \rf;9;, there is 
no $-1$ range. The prefactor in \rf;19; gives the consistency property 
$Q_e(-\xi_0) = O(1/\xi_0)$ if the $-\al$ range is extrapolated toward the 
central peak of $Q_e$.

Substitution of \rf;19; into \rf;16; yields
\opn
\int_{-\infty}^{-\xi_\al} \xi_eF_e(\xi_e)d\xi_e = 
(\xi_0/\xi_\al)^{\al-3}O(\xi_0^2).
\qno;20;

If $\al>3$, the dissipation measured by $\int_{-\xi_\al}^\infty 
\xi_eF_e(\xi_e)d\xi_e$ equals the total dissipation of $\xi_e$ in the limit 
$R\to\infty$. If $\al=3$, \rf;20; shows that there is additionally an 
essential contribution $\int_{-\infty}^{-\xi_\al} \xi_eF_e(\xi_e)d\xi_e$ that 
is $O(\xi_0^2)$ in the limit. This arises from $O(\xi_0^2/\xi_S^2)$ levels of 
$F_e(\xi_e)$ needed to induce the fast fall-off of $Q_e(\xi_e)$ at $|\xi_e| = 
O(\xi_S)$. The fast fall off occurs also if $\al > 3$, but in that case 
\rf;20; shows that the associated contribution $\int_{-\infty}^{-\xi_\al} 
\xi_eF_e(\xi_e)d\xi_e$ is a vanishing part of the overall budget in the 
$R\to\infty$ limit.

In \cite{7}, the limit ``$-\infty$'' in \rf;16; above and other equations is 
taken to mean a point within the infinite $-\al$ range, rather than true 
negative infinity ($\ll -\xi_S$). In other words ``$-\infty$'' is taken to 
mean $-\xi_\al$, as defined by \rf;18; in the limit $R=\infty$. This causes no 
problem if $\al>3$. If $\al=3$, it is clear from \rf;20; that ``$-\infty$'' 
must stay at true negative infinity.

If the possibility $\al=3$ is to be examined, clearly one cannot confine 
attention solely to the strict $-\al$ powerlaw range and ignore the transition 
region at $|\xi_e| = O(\xi_S)$. The case $\al=3$ implies that, for most fluid 
elements with steepening negative $\xi$, the inviscid steepening halts at 
$\xi$ within the transition region. The magnitude of $F_e(\xi_e)$ at $|\xi_e| 
= O(\xi_S)$ required by $\al=3$ signifies that shocks have a relatively high 
concentration in fluid with such $\xi_e$. In comparison with $\al > 3$, shocks 
are moved from environments with $|\xi_e| \ll \xi_S$ to environments in the 
transition region $|\xi_e|=O(\xi_S)$.

If an explicit shock-growth model is not adopted to fix $F_e$, it is possible 
{\it a priori} that shocks with environments in the transition region could 
fall within the picture invoked by \cite{7}, in which shocks are created at 
zero amplitude and the balance is described fully by \rf;13; and \rf;14;. In 
this case it is only needed to include the transition region of shock 
environments in $\av{\xi_e}$ when calculating the balance \rf;13; for $\al=3$. 
However, it is also possible {\it a priori} that $F_e(\xi_e)$ is weighted to 
sufficiently large $|\xi_e|$ that \rf;14; is not an accurate description of 
the dissipation mechanism for $\al=3$, or that $F_e(\xi_e)$ includes 
interactions with shocks created at finite amplitude as also considered in 
\cite{7}.

I am grateful to S. Boldyrev, W. E, U. Frisch, T. Gotoh, A. M. Polyakov, and 
E. Vanden Eijnden for fruitful interactions. This work was supported by the U. 
S. Department of Energy under Grant DE-FG03-90ER14118 and by the U. S. 
National Science Foundation under Grant DMS-9803538.

\vskip -.25in

\opbib

\vbox to -.6in { }

\bibitem[*]{0} Electronic address: rhk@lanl.gov

\bb{1} V. Yakhot and A. Chekhlov, ``Algebraic tails of probability functions 
in the random-force-driven Burgers turbulence,'' Phys. Rev. Lett. \f.77., 3118 
(1996).

\bb{2} J.-P. Bouchaud and M. M\'ezard, ``Velocity fluctuations in forced 
Burgers turbulence,'' Phys. Rev. E \f.54., 5116 (1996).

\bb{3} A. M. Polyakov, ``Turbulence without pressure,'' Phys. Rev. E \f.52., 
6183 (1995).

\bb{4} S. A. Boldyrev, ``Velocity-difference probability density functions 
for Burgers turbulence,'' Phys. Rev. E \f.55., 6907 (1997).

\bb{5} T. Gotoh and R. H. Kraichnan, Phys. Fluids \f.10., 2859 (1998).

\bb{6} W. E, K. Khanin, A, Mazel, and Y. Sinai, ``Probability distribution 
functions for the random forced Burgers equation,'' Phys. Rev. Lett. \f.78., 
1904 (1997).

\bb{7} W. E and E. Vanden Eijnden, ``Asymptotic theory for the probability 
density functions in Burgers turbulence,'' chao-dyn/9901006, submitted to 
Phys. Rev. Lett.

\bb{8} W. E and E. Vanden Eijnden, ``Another Note on Forced Burgers 
Turbulence,'' chao-dyn/990129, submitted to Phys. Fluids.

\bb{9} W. E. and E. Vanden Eijnden ``Statistical theory for the stochastic 
Burgers equation in the inviscid limit,'' chao-dyn/9904028, to appear in Comm. 
Pure App. Math.

\bb{10} A. M. Polyakov, private communication (1998)

\clbib

\end{multicols}

\end{document}